\documentclass[fleqn,usenatbib]{mnras}
\usepackage[T1]{fontenc}

\DeclareRobustCommand{\VAN}[3]{#2}
\let\VANthebibliography\thebibliography
\def\thebibliography{\DeclareRobustCommand{\VAN}[3]{##3}\VANthebibliography}

\usepackage{graphicx}	
\usepackage{amsmath}	
\usepackage{amssymb}	

\newcommand{\Msunh}{\,{\rm M}$_{\odot}$\,\,\ifmmode h^{-1}\else $h^{-1}$\fi}
\newcommand{\kms}{\,{\rm km s}\ifmmode ^{-1}\,\else $^{-1}$\,\fi}
\newcommand{\Mpch}{\,{\rm Mpc}\,\ifmmode h^{-1}\else $h^{-1}$\fi}
\newcommand{\kpch}{\,{\rm kpc}\,\ifmmode h^{-1}\else $h^{-1}$\fi}

\title[Cosmic Web Graph Entropy]{The cosmic web through the lens of graph entropy}

\author[Garc\'ia-Alvarado et al.]{
M. V. Garc\'ia-Alvarado,$^{1}$\thanks{E-mail: mv.garcia@uniandes.edu.co}
X.-D. Li$^{2}$
and J. E. Forero-Romero$^{1}$\thanks{E-mail: je.forero@uniandes.edu.co}
\\
$^{1}$Departamento de F\'isica, Universidad de los Andes, Cra. 1 No. 18A-10 Edificio Ip, CP 111711, Bogot\'a, Colombia\\
$^{2}$School of Physics and Astronomy, Sun Yat-Sen University, Guangzhou 510297, P.R.China\\
}

\date{Accepted XXX. Received YYY; in original form ZZZ}

\pubyear{2020}

\begin{document}
\label{firstpage}
\pagerange{\pageref{firstpage}--\pageref{lastpage}}
\maketitle

\begin{abstract}
  We explore the information theory entropy of a graph as a scalar to
  quantify the cosmic web. 
  We find entropy values in the range between 1.5 and 3.2 bits. We argue that this entropy can be
  used as a discrete analogue of scalars used to quantify the connectivity in continuous density fields. 
  After showing that the entropy clearly distinguishes between clustered and random points, 
  we use simulations to gauge the influence of survey geometry, cosmic variance, 
  redshift space distortions, redshift evolution, cosmological parameters and 
  spatial number density.  
  Cosmic variance shows the least important influence while
  changes from the survey geometry, redshift space distortions, cosmological 
  parameters and redshift evolution produce larger changes on the
  order of $10^{-2}$ bits.
  The largest influence on the graph entropy comes from changes in the 
  number density of clustered points.
  As the number density decreases, and the cosmic web is less pronounced, the entropy can diminish up to $0.2$ bits.
 The graph entropy is simple to compute and can be applied both to simulations and observational data from large galaxy redshift surveys; it is a new statistic that can be used in a complementary way to other kinds of topological or clustering measurements. 
\end{abstract}


\section{Introduction}

The cosmic web is one of the most salient features of the matter distribution
on cosmological scales. 
There is a great variety of methods striving to locally describe this complex 
structure by classifying a region as belonging either
to a void, filament, wall or cluster \citep{2018MNRAS.473.1195L}. 

Global descriptors are also useful to measure the topological structure of this web.
Quantities such as the topological invariants described by  the genus
\citep{1986ApJ...309....1H, 1986ApJ...306..341G} the Minkowski functionals 
\citep{1997ApJ...482L...1S}, the Betti numbers
\citep{2013JKAS...46..125P,2017MNRAS.465.4281P}, {or the average
number of filaments connected to a cluster \citep{2018MNRAS.479..973C}, have been widely applied to describe the cosmic web. 

Another scalar, the information theory entropy based on the inhomogeneous 
spatial matter density in the cosmic web,
$S=-\int\rho(\bf{r})\log\rho(\bf{r}) d\bf{r}$, has
been used to derive analytical expressions for its cosmological evolution
\citep{2004PhRvL..92n1302H}, measurements from surveys
\citep{2015MNRAS.454.2647P}
and estimates from simulations \citep{2020MNRAS.491.5447V}.

The same information theory principles can be extended to measure the entropy built 
on graphs  used to describe the cosmic web. 
This entropy can be computed from the number of connections for each
point in the graph,  which is a property that  can be computed for commonly used graphs in
this context, such as  minimal spanning trees
\citep{1985MNRAS.216...17B}, Delaunay tessellations 
\citep{2007MNRAS.382....2R}, neighbor networks within a fixed linking
length \citep{2016MNRAS.459.2690H} or the $\beta$-skeleton
\citep{2019MNRAS.485.5276F}.   
For any of those graphs one can estimate the probability of a point
having $n$ connections, $P_n$, and from these values a global entropy
can be defined as $S = \sum_{P_n>0}-P_n\log_2{P_n}$. 

The purpose of this Letter is to advocate the information theory entropy on a graph
as a global scalar to describe the cosmic web.
We use the $\beta$-skeleton graph to explore a whole graph family 
by varying the real valued parameter $\beta$.
As cosmic web tracers we use dark matter halos from cosmological N-body simulations. 
This Letter is structured as follows. 
In Section 2 we present the $\beta$-skeleton and the graph entropy definition.
In Section 3 we summarize the main features of the cosmological simulations we use and 
the basic setup of the numerical experiments we want to run.
The results are presented in Section 4.
We discuss the interpretation of our findings in Section
5 to finally summarize our Conclusions in Section 6.  

\section{Graphs and Entropy}

\subsection{The $\beta$-Skeleton Graph}

The $\beta$-Skeleton is a non-directed graph that determines the
connectivity between pairs of points as a function of the real
parameter $\beta$ \citep{1985Kirkpatrick}. 
A pair of points, $p$ and $q$, are connected by an edge if the exclusion region parameterized by $\beta$ does not include any other third point.
We use the so-called \emph{lune} definition for this exclusion region.
For $\beta\geq 1$ the lune is defined as the intersection of two congruent spheres with diameter $D= \beta d$, with $d$ being the distance between $p$ and $q$.
The spheres are centered on positions $\pm d(1-\beta)/2$, measured from the midpoint between $p$ and $q$, lying along the line connecting the two points.
Figure \ref{fig:example} illustrates how the exclusion regions change for the range of $\beta$ values explored in this Letter.

Under this definition the volume of the exclusion region is a
monotonic function of $\beta$.
As the exclusion regions grows it is less likely to find pairs that are connected \citep{adamatzky2013growing}.
The $\beta$-skeleton is a connected graph for $1\leq \beta\leq 2$
(i.e. all nodes have at least
one connection) and could be a disconnected graph for $\beta>2$ \citep{bose2002spanning}.
More details on the applications of this graph on large scale structure data
can be found in \citep{2019MNRAS.485.5276F}.

\subsection{Graph Entropy Definition}

After the graph has been constructed it is possible to estimate $P_n$, the probability of 
a node to have $n$ connections.
From these values we define the graph entropy as

\begin{equation}
\centering
    S = \displaystyle\sum_{P_n>0}-P_n\log_2{P_n}.
    \label{ecuacionEntropia}
\end{equation}
This graph entropy definition is the simplest possible as it is global and only 
takes into account the degree of connectivity \citep{2012Entrp..14..559M}. 
Figure \ref{fig:probabilities} shows some $P_n$ distributions for the 
$\beta$-skeleton graphs built on data from an N-body simulation.

\begin{figure}
    \centering
    \includegraphics[width=0.45\textwidth]{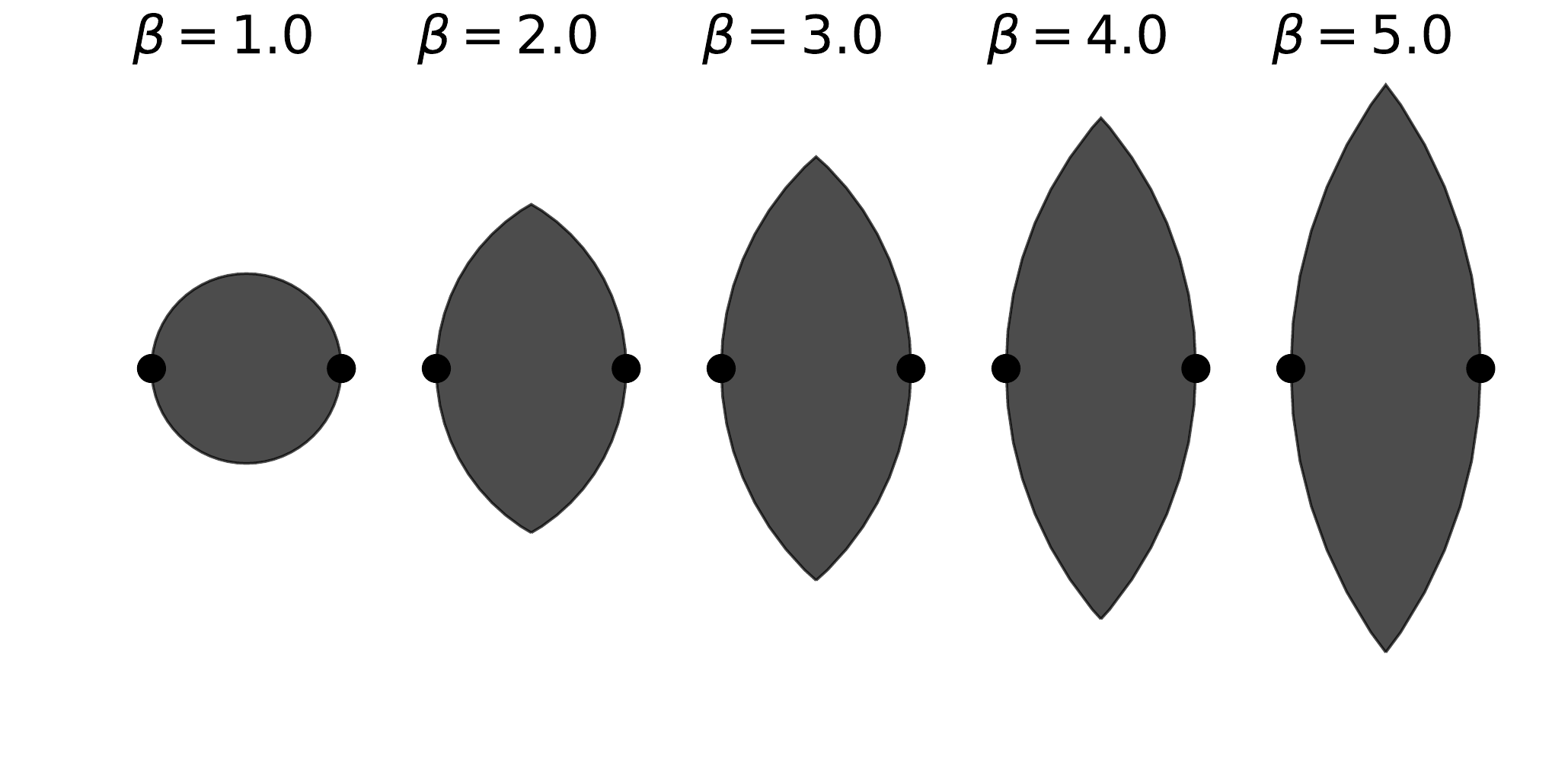}
   \caption{Illustration of the of exclusion regions (shaded areas) between two points using the lune-based definition for five different values of $\beta$ explored in this Letter.
 \label{fig:example}}  
\end{figure}

\begin{figure}
    \includegraphics[width=0.45\textwidth]{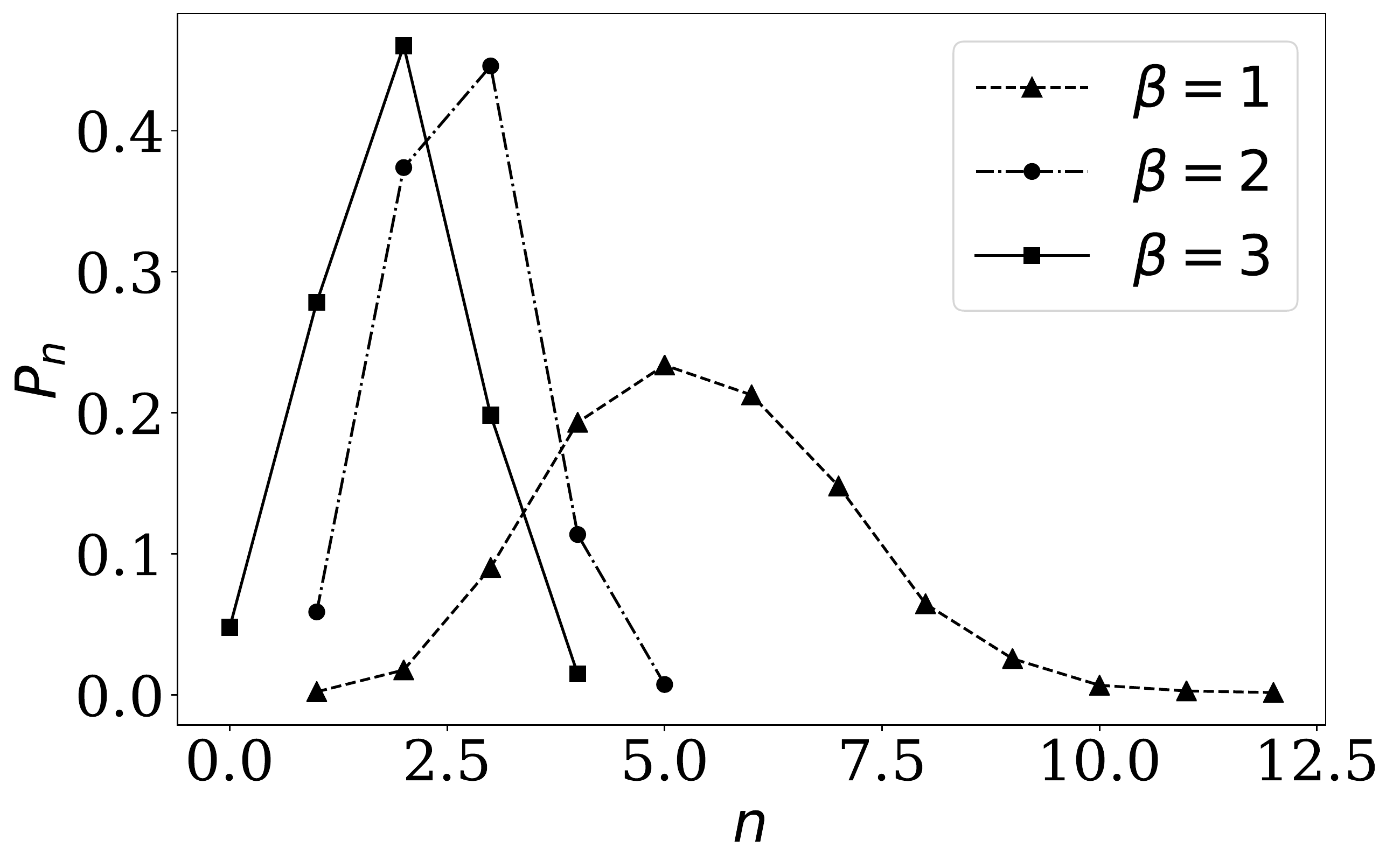}
    \caption{Probabilities of having $n$ connections, $P_n$, for three different values of
    $\beta$. 
    The graph entropy summarizes the changes in the $P_n$ distribution as a function of $\beta$.}
    \label{fig:probabilities}
\end{figure}

\section{Methods}

\subsection{Simulations and Mock Catalogs}

We use data products from the Abacus project \citep{abacus}.
That project is a set of dark matter only N-body simulations that follows
the growth of structure in an explicit cosmological setup.
It includes simulations with different box sizes and cosmological parameters.

Here we focus on the simulations done in a cubic box of $720$\Mpch on a side
with $1440^3$ particles. 
This resolution corresponds to a particle mass of $\sim1\times10^{10}$ \Msunh.
The same box was run with $20$ different initial conditions at a fixed value for
the cosmological parameters and also with fixed initial conditions and $40$ different
sets of cosmological parameters.

We use the Friend-of-Friends catalogs to build all the datasets to be used
as an input to the $\beta$-skeleton. In all cases we take only into account halos
with a maximum circular velocity larger than $300$\kms. 
We then apply different geometrical cuts and changes to their positions to build 
our final mock catalogs.

We build mock catalogs with two different geometries: spheres and spherical shells.
Both geometries have a maximum radius of $300$ \Mpch.
The shells have an inner radius of $250$\Mpch.
For each one of these geometries we build the corresponding random catalogs
by fixing their radial coordinate with respect to the geometrical center and 
randomize its angular coordinates.
We also produced catalogs with and without Redshift Space Distortion (RSD) effects. 
To study the effect of redshift evolution we use spheres extracted from
the snapshots at redshifts of $z=0.1$, $0.3$, $0.5$, $0.7$, $1$ and $1.5$.

\subsection{Numerical Experiments}

For each mock catalog we build the $\beta$-Skeleton and estimate each $P_n$ as  $\frac{N_e}{T_n}$, where $N_e$ is the number of nodes with $n$ connections and $N$ is the
total number of nodes in the data set.
From this set of $P_n$ values we measure the graph entropy using Equation \ref{ecuacionEntropia}.

We only use the values $\beta=1.0$, $1.5$, $2.0$, $2.5$, $3.0$, $3.5$, $4.0$, $4.5$ and $5$.
In what follows we preset results as a function of $\beta$ by continous lines.
We have checked that including more $\beta$ values does not significantly change
our results. 
In other words the graph entropy turns out to be a slowly varying function of $\beta$.

Our experiments focus on quantifying the effects on the entropy from:
cosmic variance, mock geometry, RSD, redshift evolution, cosmological
parameters and spatial number density.

\begin{figure}
    \includegraphics[width=0.45\textwidth]{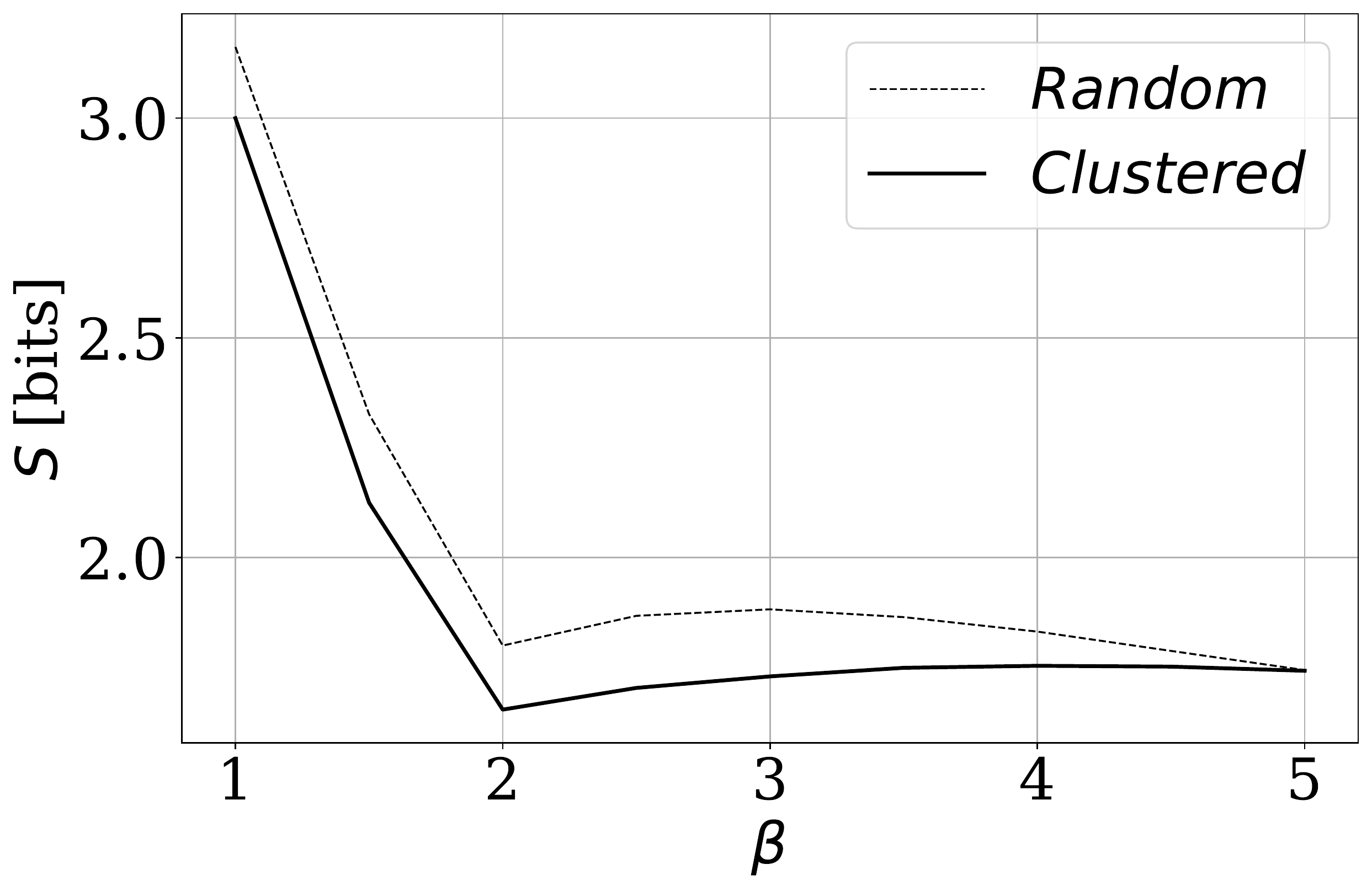}
    \caption{Graph entropy as a function of the parameter $\beta$ 
    for clustered and random points distributed inside a sphere. \label{fig:entropy}}
\end{figure}

\section{Results}
\label{sec:results}

\begin{figure*}
    \centering
    \includegraphics[width=0.38\textwidth]{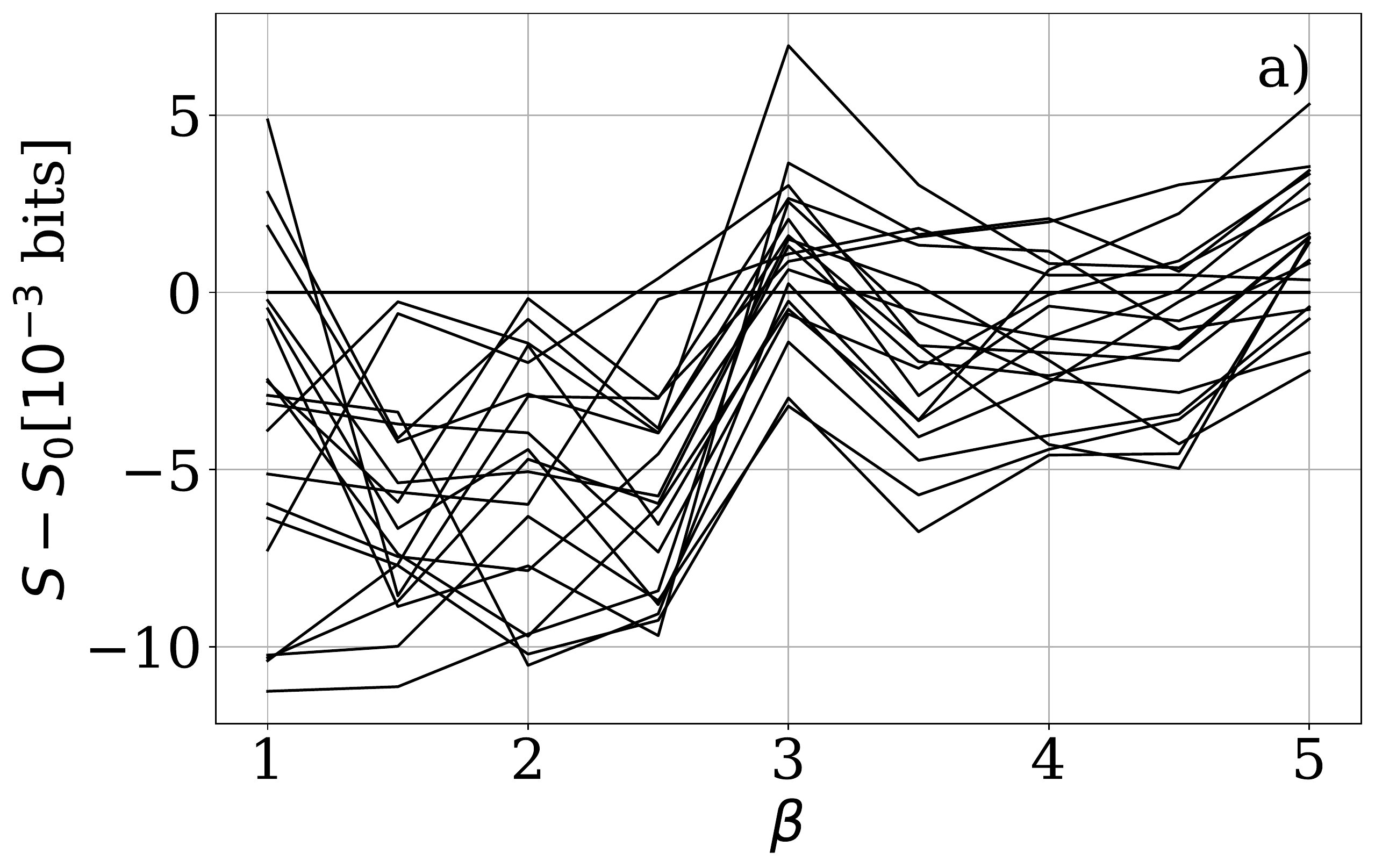} \hspace{0.5cm}
    \includegraphics[width=0.38\textwidth]{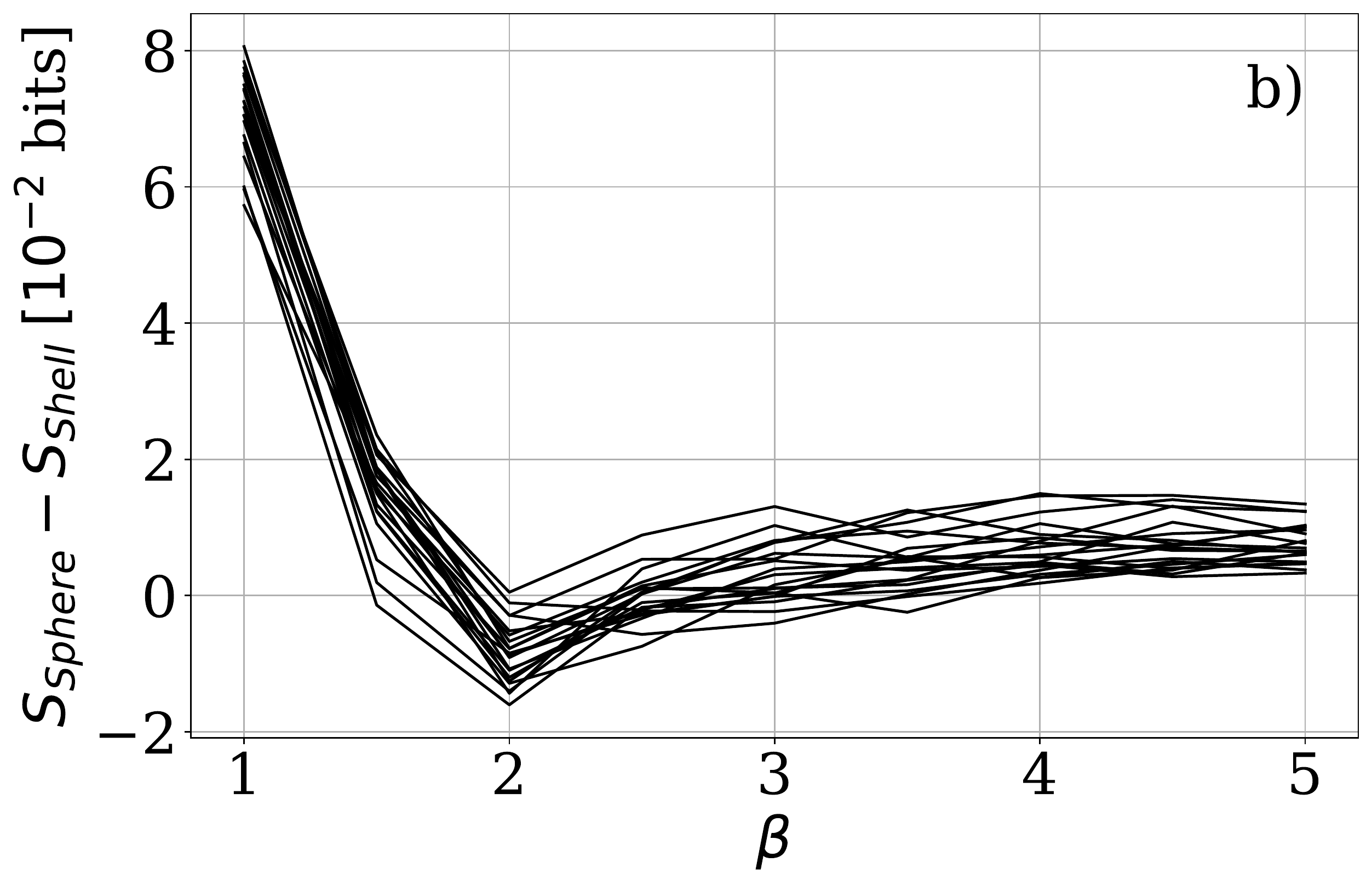} 
    \includegraphics[width=0.38\textwidth]{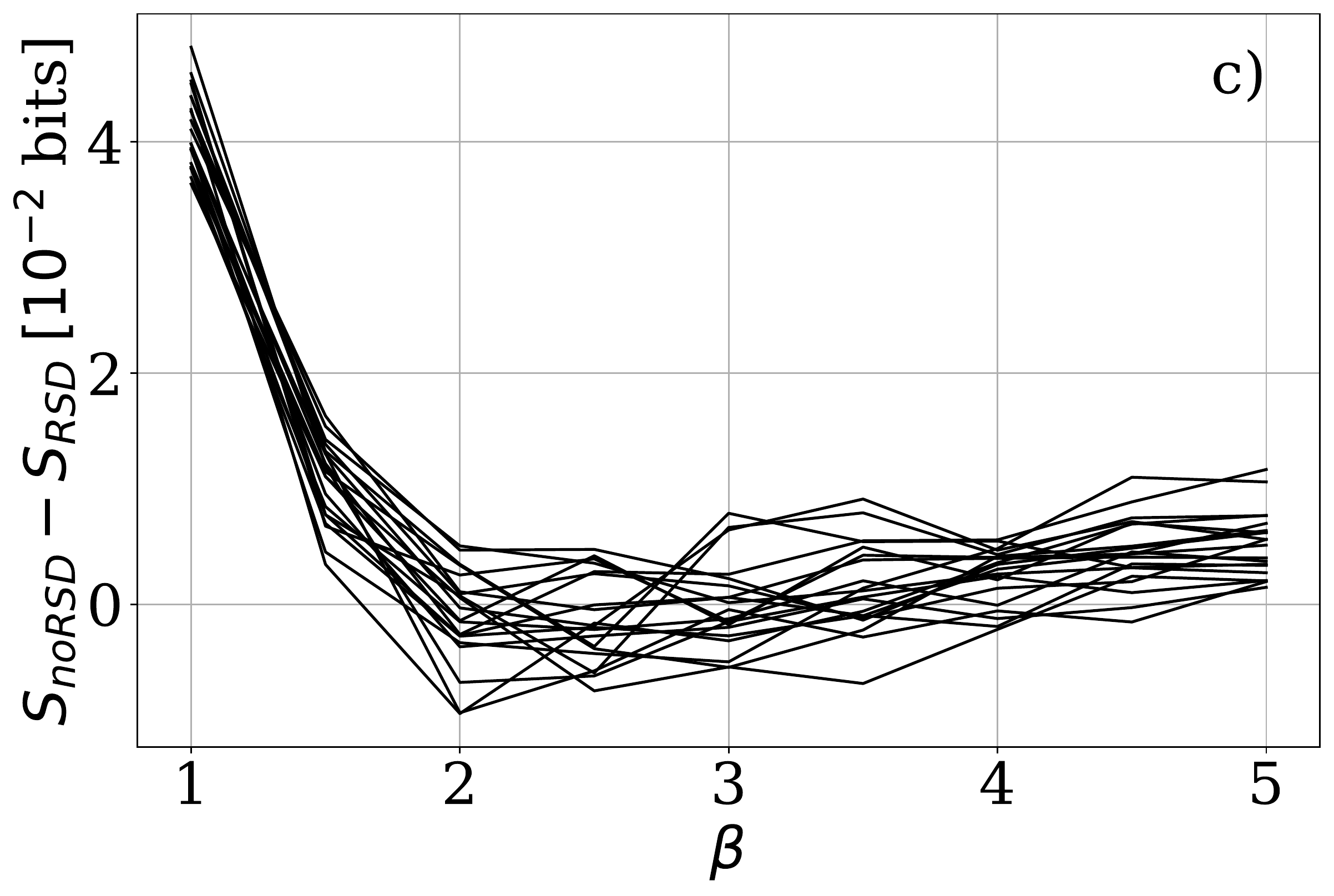} 
    \hspace{0.5cm}
    \includegraphics[width=0.38\textwidth]{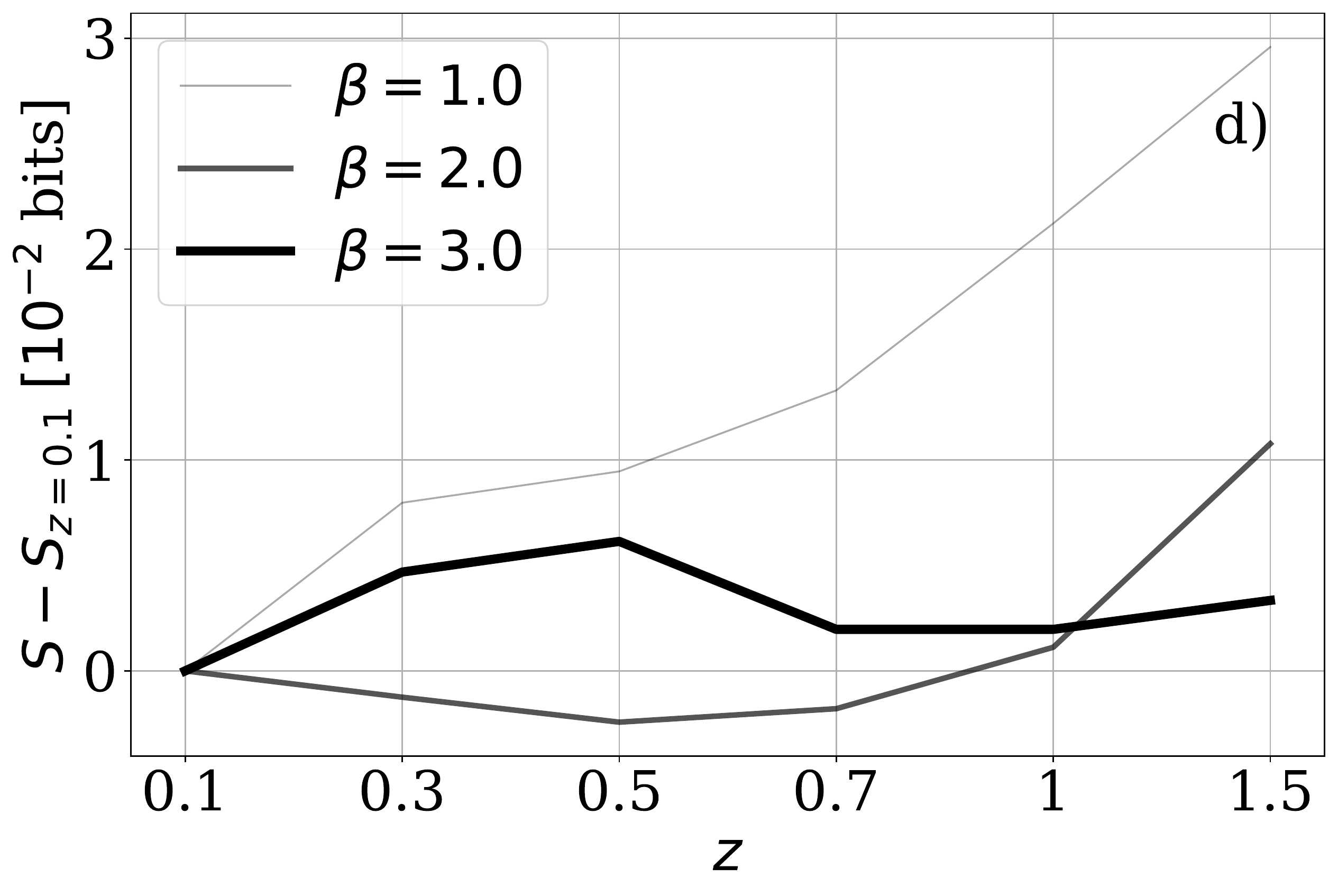} 
    \includegraphics[width=0.38\textwidth]{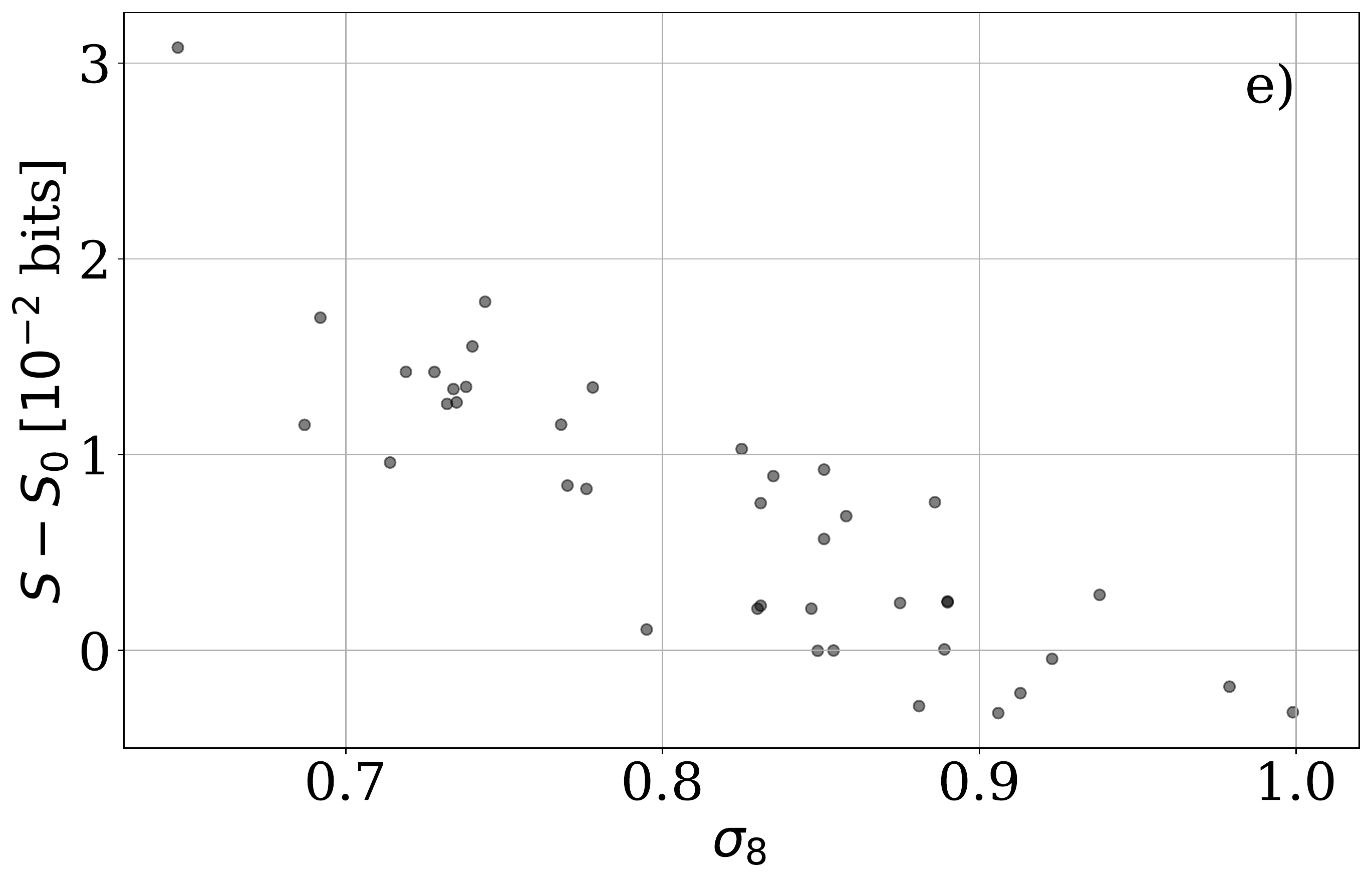}
    \hspace{0.5cm}
    \includegraphics[width=0.38\textwidth]{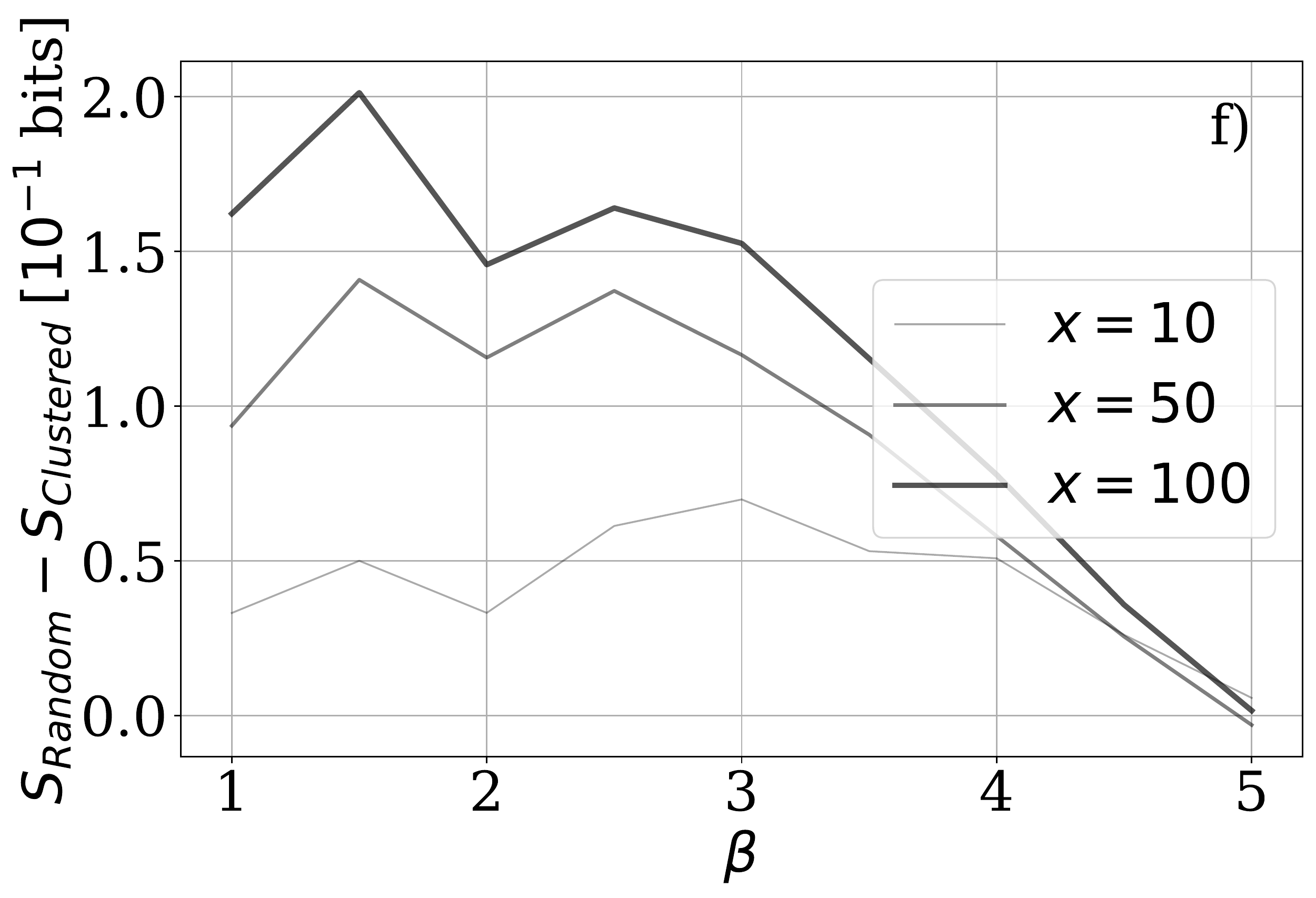}
    \caption{Influence of different effects on the graph entropy. In all cases we 
    report the changes in entropy from a reference value. 
    Details are explained in the Results section. $a)$ Cosmic variance. 
    $b)$ Survey geometry. 
    $c)$ Redshift Space Distortions (RSD).
    $d)$ Redshift evolution.
    $e)$ Cosmological parameters. We only show the results for $\sigma_8$, 
    the parameter that shows the strongest correlation with the graph entropy.
    The plot corresponds to the entropy computed for the $1$-skeleton at $z=0.1$. 
    $f)$ Number density. 
    This calculation is performed on $20$ different spheres at $z=0.1$.
    A different percentage (shown in the caption) of these points are sampled. 
    We show the difference with respect to the entropy measured on the spheres 
    of sampled random points.\label{fig:diferencias}}
\end{figure*}

Figure \ref{fig:entropy} shows the graph entropy as a function of $\beta$ for
clustered and random points distributed over a sphere.
The most important message from this plot is that the entropy can distinguish between
clustered and random points.

This result immediately follows from the fact that the connectivity distribution of clustered and random points is different. 
$P_n$ follows a narrower distribution for random points than it does for clustered points.

Figure \ref{fig:entropy} also shows that the entropy follows a decreasing trend with increasing $\beta$ that seems to be interrupted when $\beta=2$, which represents a local minimum of the curve.
The steep decreasing trend for $\beta<2$ is a direct consequence of the monotonic and strong decrease in the number of connections as $\beta$ increases.
Figure \ref{fig:probabilities} shows how the connection distribution becomes significantly narrower as $\beta$ goes from $1$ to $2$, explaining the entropy decrease.

For $\beta>2$ the same connectivity distribution does not become significantly narrower (it is hard to get points with much less than $n=4$ connections).
At the same time, at $\beta=2$ the constructed graph goes from
being fully connected ($P_0=0$) to have some isolated nodes ($P_0>0$) for $\beta>2$.
These two facts (the impossibility of having a much narrower distribution and the availability of the new state $n=0$) account for the slow increase in the entropy for $\beta>2$.
\indent
Another interesting point is that around $\beta=5$ the two entropies
are equal.  
This corresponds to the extreme of a very sparse graph where the
clustered and random point distributions become indistinguishable from the point of view of the entropy.

The general shape shown in Figure \ref{fig:entropy} is conserved in
all cases.  
In what follows we quantify the entropy changes under different
conditions.

\subsection{Cosmic Variance}

We measure the effects of cosmic variance by comparing the entropy of the spherical 
mocks built from the $20$ different realizations of the standard
cosmology at $z=0.1$ without considering the effects of RSD.  
Panel $a)$ in Figure \ref{fig:diferencias} shows the entropy
difference with respect to a reference mock. The variance of the
entropy differences is on the order of $10^{-3}$ bits, almost
independent of $\beta$. 
The same result holds for mocks with shell geometry.

\subsection{Geometry}

We evaluate geometry influence by comparing spheres and shells.
We compare each spherical mock with its corresponding shell.
This is done at $z=0.1$ without RSD effects over the $20$ different realizations
of the standard cosmology.
Panel $b)$ in Figure \ref{fig:diferencias} shows that the entropy
difference has a strong dependence with $\beta$.  
For $\beta=1$ points on the spherical distribution have $8\times
10^{-2}$ more bits of entropy than shells. 
For values around $\beta=2$ this difference drops to negative values
(shells have more entropy than spheres) but only on the order of
$1\times10^{-2}$ bits.

\subsection{Redshift Space Distortions}

We measured the RSD influence on the spheres extracted from the $20$ different realizations of the standard cosmology at $z=0.1$.
In the mock building process the RSD effect is applied before making the geometrical cut.
Panel $c)$ in Figure \ref{fig:diferencias} shows a similar picture than the one found in the previous subsection.
Namely, a strong dependence for $\beta\leq 2$, with more entropy in the data without RSD and a flat response for $\beta\geq 2$.
In this case the differences continue to be on the order of $10^{-2}$ bits, with the
largest difference located at $\beta=1$.

\subsection{Redshift Evolution}
We use six spheres to measure the redshift evolution of the graph entropy. 
Each sphere comes from a simulation with the standard cosmology with redshifts of
$z=0.1$, $0.3$, $0.5$, $0.7$, $1.0$ and $1.5$. 
We used the spheres in comoving coordinates without taking into account any form of RSD.
Panel $d)$ in Figure \ref{fig:diferencias} shows the entropy evolution computed for values
of $\beta=1$, $2$ and $3$.
For each $\beta$ we show the entropy differences with respect to the entropy at $z=0.1$.
Each $\beta$ value produces a different entropy redshift evolution.
The largest variation is $3\times 10^{-2}$ bits for the $1$-skeleton between the redshift of
$z=1.5$ and $z=0.1$, where the entropy decreases monotonously with decreasing redshift.
This redshift evolution is different for the $2$-skeleton and the $3$-skeleton; the changes 
are not monotonous and less pronounced.

\subsection{Cosmological Parameters}
The Abacus simulations have $40$ different realizations with different values for the
cosmological  parameters $H_0$, $\Omega_{DE}$, $\Omega_{M}$, $n_s$, $\sigma_8$ and $w_0$.
We measure the entropy differences with respect to the entropy for the cosmology of 
reference in the $40$ different spheres at $z=0.1$, without RSD effects, 
for the skeletons with $\beta=1$, $2$ and $3$.
Panel $e)$ in Figure \ref{fig:diferencias} 
shows the entropy differences as a function of $\sigma_8$ for $\beta=1$.
Increasing values of $\sigma_8$ correspond to lower entropy values. 
This is the strongest correlation we find between entropy and a cosmological parameter.
For $\beta=2$ the correlation is weaker and for $\beta=3$ the correlation inverses its
trend.
The different cosmological parameters induce changes on the entropy of magnitude $3\times10^{-2}$ bits at most.

\subsection{Number densities}

We prepare and additional dataset where we randomly 
sample a percentage (between $10\%$ and $100\%$ in jumps of $10\%$) 
of the points from the original random and clustered spheres.
Although the number density changes, the different points have the 
same correlation function as they are dominated by dark matter halos with circular velocities $\approx 300$ \kms.

The entropy changes in this case are the largest among our series of experiments.
Panel $e)$ in Figure \ref{fig:diferencias} shows the entropy differences 
with respect to the corresponding random sphere as a function
of $\beta$.
The entropy differences decrease with the number density. 
The differences range between $0.05$ to $0.2$ bits.

A central result is that the entropy only changes significantly in the
clustered points. 
If we compute the entropy on the spheres with random distribution of
points  with varying number densities, then the entropy changes among
those spheres  is only on the order of $10^{-2}$ bits. 

\section{Discussion}

From all these findings, how do we interpret the entropy measured on the graphs built from 3D points?
Let us recall that the graph encodes topological
features from the data.
  Its entropy focuses on a single aspect: the connectivity distribution.

In a graph, regions with larger number of connections tend to be in regions with high number density.
It means that the connectivity distribution traces to some extent the point density distribution.
  The graph entropy can be thought as a summary
  statistic of the point density distribution.

  For the $\beta$-skeleton graph, as $\beta$ increases, only the points in the densest regions remain
  connected \citep{2019MNRAS.485.5276F}.
  This means that the entropy as a function of $\beta$ describes the
  balance between the number of 
  overdense regions in the point distribution (which are still
  connected) and the low density regions (that will be disconnected
  for large values of $\beta$.)

As an additional test of this reasoning we computed the Voronoi tesellation over the datasets used in this Letter. 
We find that the shape of the distribution for the Voronoi cell volumes (a measure
of point density; large volumes can be interpreted as low density regions) 
has similar properties as the graph entropy: there is clear difference between clustered and random points, it is invariant for random points,
weakly dependent on cosmic variance, geometry, RSD, redshift and
cosmological parameters and strongly dependent on the number density
of clustered points.
Furthermore, as $\beta$ increases, the cell volume distributions for the points included in
the graph selects mostly regions of large density (small volumes.)

A complementary point of view is to consider that the local density
correlates with cosmic web features,
i.e. cluster-like regions are overdense, voids are underdense, while filaments and sheets have
 intermediate densities \citep{2018MNRAS.473.1195L}. 
 Changes in the number density clearly impact the cosmic web defined by the tracers.
  As we sample the input data and the number density decreases, we observe how the voids become
  larger and the filamentary patterns are less pronounced.  
  To quantify this visual impression we estimated the typical void size
  (normalized by the mean interparticle distance)
  in the datasets with changing number density.  
We find that for clustered points larger entropy values correspond to smaller voids and for random points the void size remains the same, as expected.\\
\indent
From this discussion we argue that the graph entropy as a function of $\beta$ can be seen as the discrete analogue of
  measuring the connectivity 
  over a continuous density field with varying density thresholds \citep{2013JKAS...46..125P}.

\section{Conclusions}

In this Letter we presented the graph entropy as a new scalar to 
quantify the large scale structure of the Universe.
The entropy definition we use is a global quantity based on 
the concept commonly used in information theory.
It uses the probability of having $n$ connections, $P_n$, to build the 
scalar $S=\sum_{P_n>0} -P_n  \log_2 P_n$. 
We based our quantitative analysis on the $\beta$-skeleton graph 
built on mock catalogs constructed from cosmological N-body simulations.

We measured that the graph entropy ranges 
between $1.5$ and $3.2$ bits.
We found that it can clearly distinguish between random and clustered points.
Then we tested the influence on the graph entropy of six major factors: 
cosmic variance, survey geometry, redshift space distortions, redshift evolution, cosmological parameters and number densities.
We found that the strongest influence on the entropy appears in
clustered data at different number densities.

The graph entropy can be used as the discrete analogue
  of measuring the connectivity in a continuous density field.
  As such, it represents a new statistic that can be used in a complementary way to other kinds of topological measurements.
 
 Further tests of the applicability to observational data and other graphs are necessary beyond the results presented here.
  An incomplete list of experiments that might be performed in the context of cosmology are:  comparing the entropy for mock and observed galaxy survey data to constrain bias models that appear indistinguishable using other statistics; measuring the entropy from different parts of the sky to check for isotropy; 
building graphs from  other quantities (i.e. features in the Cosmic Microwave Background, weak lensing peaks,
basin points from reconstructed peculiar velocity fields) to measure its entropy and quantify its connectivity properties.

\section*{Acknowledgements}
XDL acknowledges the support from NSFC grant (No. 11803094).

\section*{Data Availability}
The datasets were derived from sources in the public domain: AbacusCosmos project, \url{https://lgarrison.github.io/AbacusCosmos/}.
The data underlying this article are available in the Beta-Skeleton-Entropy repository at \url{https://github.com/mvgarcia/Beta-Skeleton-Entropy}

\bibliographystyle{mnras}

\end{document}